\documentstyle[12pt,epsf,epsfig,wrapfig]{article}
\begin{document}%
\begin{center}%
 {\Large\bfseries RELATIONS BETWEEN GPDS AND TMDS:\\[1ex]MODEL RESULTS AND BEYOND}%
 \vskip 5mm%
 S. Mei{\ss}ner$^{1}$, A. Metz$^{2}$, M. Schlegel$^{3}$, and K. Goeke$^{1}$%
 \vskip 5mm%
 {\small(1) {\it Institut f{\"u}r Theoretische Physik II, Ruhr-Universit{\"a}t Bochum, 44780 Bochum, Germany}\\%
 (2) {\it Department of Physics, Temple University, Philadelphia, PA 19122-6082, USA}\\%
 (3) {\it Theory Center, Jefferson Lab, 12000 Jefferson Avenue, Newport News, VA 23606, USA}}%
\end{center}%
\vskip 5mm%
\begin{abstract}%
Comparing the respective structures of the correlators defining generalized and transverse momentum dependent parton distributions, one finds possible relations between these two objects. Although it looks like the relations found do not hold in general, we show that they can be established at least in simple spectator model calculations. In addition, we discuss these relations in the context of generalized transverse momentum dependent parton distributions.
\end{abstract}%
\vskip 8mm%
\section{Introduction and definitions}
Parton distributions are an essential tool for the QCD-description of hadronic scattering processes. In particular, generalized parton distributions (GPDs) and transverse momentum dependent parton distributions (TMDs), which appear in connection with hard exclusive and semi-inclusive reactions, respectively, attracted a lot of interest during the last years. Although these two types of parton distributions are \emph{a priori} two distinct objects, recent work suggests possible relations between them~\cite{Burkardt:2002ks:2003uw, Burkardt:2003je, Diehl:2005jf, Burkardt:2005hp, Lu:2006kt, Meissner:2007rx}. In this note, we will briefly summarize the current knowledge on these relations (as presented in the review article~\cite{Meissner:2007rx}) and, in addition, also present new results from the analysis of generalized transverse momentum dependent parton distributions (GTMDs).

Before discussing the relations between GPDs and TMDs we recall their definitions. First of all, the leading twist GPDs of the nucleon for unpolarized quarks are defined through
\begin{eqnarray}
 F^q(x,\xi,\vec{\Delta}_T;\lambda,\lambda')
 &\!\!\!=\!\!\!&\frac{1}{2}\int\frac{dz^-}{2\pi}\,e^{i k\cdot z}\,\big<p';\lambda'\big|\,
  \bar\psi\big(\!-\!{\textstyle\frac{1}{2}}z\big)\,\gamma^+\,{\mathcal{W}}_{\mathrm{GPD}}\,
  \psi\big({\textstyle\frac{1}{2}}z\big)\,\big|p;\lambda\big>\,\Big|_{z^+=\vec{z}_T=0} \nonumber\\
 &\!\!\!=\!\!\!&\frac{1}{2P^+}\,\bar u(p',\lambda')\,\bigg(\gamma^+\,H^{q}(x,\xi,t)
  +\frac{i\sigma^{+\mu}\Delta_\mu}{2M}\,E^{q}(x,\xi,t)\bigg)\,u(p,\lambda)\,,
  \label{eq:DefGPDs}
\end{eqnarray}
with the average nucleon momentum $P=\textstyle\frac{1}{2}(p+p')$ and the nucleon momentum transfer $\Delta=p'-p$. The GPDs depend on the three kinematical variables
\begin{equation}
 x=\frac{k^+}{P^+} \,, \qquad
 \xi=-\frac{\Delta^+}{2P^+} \,, \qquad
 t=\Delta^2 \,.
\end{equation}
Throughout this note, we disregard any dependence of the correlator in Eq.~(\ref{eq:DefGPDs}) on a renormalization scale $\mu$, as this does not affect the analysis of possible relations between GPDs and TMDs.

To derive these relations, it is convenient to work with the GPDs in impact parameter instead of momentum space. This representation of the GPDs can be obtained by Fourier transforming the correlator in Eq.~(\ref{eq:DefGPDs}) for $\xi=0$,
\begin{equation}
 {\mathcal{F}}^q(x,\vec{b}_T;S)
 =\int\frac{d^2\vec{\Delta}_T}{(2\pi)^2}\,e^{-i\vec{\Delta}_T\cdot\vec{b}_T}\,
  F^q(x,0,\vec{\Delta}_T;S)
 ={\mathcal{H}}^q(x,\vec{b}_T^{\,2})
  +\frac{\epsilon_T^{ij}b_T^iS_T^j}{M}\,\bigg({\mathcal{E}}^q(x,\vec{b}_T^{\,2})\bigg)'\,,
 \label{eq:GPDsIPS}
\end{equation}
where $S$ parametrizes all possible combinations of the helicities $\lambda$ and $\lambda'$ as described in Refs.~\cite{Diehl:2005jf, Meissner:2007rx}. The GPDs ${\mathcal{H}}^q$ and ${\mathcal{E}}^q$ are the Fourier transformed GPDs $H^q$ and $E^q$, respectively, and the prime denotes the first derivative with respect to $\vec{b}_T^{\,2}$. The correlator in Eq.~(\ref{eq:GPDsIPS}) can be interpreted as the probability density of finding an unpolarized quark with longitudinal momentum fraction $x$ at transverse position $\vec{b}_T$ inside a transversely polarized nucleon.

The second set of parton distributions we are interested in are the TMDs. The leading twist TMDs of a nucleon for unpolarized quarks are defined through
\begin{eqnarray}
 \Phi^q(x,\vec{k}_T;S)
 &\!\!\!=\!\!\!&\frac{1}{2}\int\frac{dz^-}{2\pi}\,\frac{d^2\vec{z}_T}{(2\pi)^2}\,e^{i k\cdot z}\,
  \big<P;S\big|\,\bar\psi\big(\!-\!{\textstyle\frac{1}{2}}z\big)\,\gamma^+\,{\mathcal{W}}_{\mathrm{TMD}}\,
  \psi\big({\textstyle\frac{1}{2}}z\big)\,\big|P;S\big>\,\Big|_{z^+=0^+} \nonumber\\
 &\!\!\!=\!\!\!&f_1^{q}(x,\vec{k}_T^{\,2})
  -\frac{\epsilon^{ij}_T k_T^i S_T^j}{M}\,f_{1T}^{\bot q}(x,\vec{k}_T^{\,2})\,,
  \label{eq:DefTMDs}
\end{eqnarray}
where again we disregard any dependence on a renormalization scale $\mu$. Similar to the GPDs in impact parameter space the TMDs have a probability interpretation, too. The correlator in Eq.~(\ref{eq:DefTMDs}) gives the probability of finding an unpolarized quark with longitudinal momentum fraction $x$ and transverse momentum $\vec{k}_T$ inside a transversely polarized target.

\section{Relations between GPDs and TMDs}
Comparing the respective structures of the correlators in Eqs.~(\ref{eq:GPDsIPS}) and~(\ref{eq:DefTMDs}) one finds that they are identical after exchanging the impact parameter $\vec{b}_T$ and the transverse parton momentum $\vec{k}_T$. This, together with the similar probability interpretations of the correlators, leads to the assumption that there might exist some relations between these two objects.

Performing such a comparison for all leading twist parton distributions for quarks~\cite{Diehl:2005jf} as well as for gluons~\cite{Meissner:2007rx}, one finds the following set of possible relations, which can be grouped into four different types according to the number of derivatives on the GPD side:
\begin{eqnarray}
 {\mathcal{H}}^{q/g}\,\leftrightarrow\ f_1^{q/g} \,,
  \ &&\ \tilde{\mathcal{H}}^{q/g}\,\leftrightarrow\ g_{1L}^{q/g} \,, \nonumber\\
 \Big({\mathcal{H}}_T^{q}-{\textstyle\frac{\vec{b}_T^{\,2}}{M^2}}\,\Delta_b\tilde{\mathcal{H}}_T^{q}\Big) \,,
  &\!\!\leftrightarrow\!\!&
  \Big(h_{1T}^{q}+{\textstyle\frac{\vec{k}_T^{\,2}}{2M^2}}\,h_{1T}^{\bot q}\Big)
  \label{eq:Rel1} \\[2ex]
 \Big({\mathcal{E}}^{q/g}\Big)'\,\leftrightarrow\ -f_{1T}^{\bot q/g} \,,
  \ &&\ \Big({\mathcal{E}}_T^{q}+2\tilde{\mathcal{H}}_T^{q}\Big)'
  \,\leftrightarrow\ -h_1^{\bot q} \,, \nonumber\\
 \Big({\mathcal{H}}_T^g-{\textstyle\frac{\vec{b}_T^{\,2}}{M^2}}\,\Delta_b\tilde{\mathcal{H}}_T^g\Big)'
  &\!\!\leftrightarrow\!\!&
  -{\textstyle\frac{1}{2}}\Big(h_{1T}^g+{\textstyle\frac{\vec{k}_T^{\,2}}{2M^2}}\,h_{1T}^{\bot g}\Big) \,,
  \label{eq:Rel2} \\[2ex]
 \Big(\tilde{\mathcal{H}}_T^{q}\Big)''\,\leftrightarrow\ {\textstyle\frac{1}{2}}h_{1T}^{\bot q} \,,
  \ &&\ \Big({\mathcal{E}}_T^g+2\tilde{\mathcal{H}}_T^g\Big)''
  \,\leftrightarrow\ {\textstyle\frac{1}{2}}h_1^{\bot g} \,,
  \label{eq:Rel3} \\[2ex]
 \Big(\tilde{\mathcal{H}}_T^g\Big)'''&\!\!\leftrightarrow\!\!&-{\textstyle\frac{1}{4}}h_{1T}^{\bot g} \,.
  \label{eq:Rel4}
\end{eqnarray}

\section{Model results\dots}
To check whether the possible relations in Eqs.~(\ref{eq:Rel1})--(\ref{eq:Rel4}) really exist, we performed model calculations in two simple spectator models: a scalar diquark spectator model of the nucleon and a quark target model in perturbative QCD. In these models we were able to confirm all relations to lowest order in perturbation theory~\cite{Meissner:2007rx}.

For the relations of first type in Eq.~(\ref{eq:Rel1}) this is not very surprising as it is a well known model-independent property of the involved GPDs and TMDs that they can be reduced to the same forward parton distributions,
\begin{equation}
 q(x)=\int d^2\vec{b}_T\,{\mathcal{H}}^q(x,\vec{b}_T^{\,2})=\int d^2\vec{k}_T\,f_1^q(x,\vec{k}_T^{\,2})
 \label{eq:ModRel1}
\end{equation}
and analogous for all other relations in Eq.~(\ref{eq:Rel1}).

In the case of the relations of second type in Eq.~(\ref{eq:Rel2}) we were able to reproduce the results of Refs.~\cite{Burkardt:2003je, Lu:2006kt} and to generalize them~\cite{Meissner:2007rx}. We suppose, however, that the explicit form for the relations in Eq.~(\ref{eq:Rel2}) presented in Refs.~\cite{Burkardt:2003je, Lu:2006kt, Meissner:2007rx} is only valid in the performed lowest order model calculations and not in general, because it will probably break down once higher order contributions are taken into account \cite{Meissner:2007rx}. Nevertheless, this type of relations has very interesting phenomenological implications~\cite{Burkardt:2002ks:2003uw, Burkardt:2003je, Diehl:2005jf, Burkardt:2005hp, Lu:2006kt, Meissner:2007rx}.

For the relations of third type in Eq.~(\ref{eq:Rel3}) we found that
\begin{equation}
 \int d^2\vec{b}_T\,\vec{b}_T^{\,2}\,\Big(\tilde{\mathcal{H}}_T^q(x,\vec{b}_T^{\,2})\Big)''
 =\int d^2\vec{k}_T\,\vec{k}_T^{\,2}\,{\textstyle\frac{1}{2}}h_{1T}^{\bot q}(x,\vec{k}_T^{\,2})
 \label{eq:ModRel3}
\end{equation}
and analogous for the other relation in Eq.~(\ref{eq:Rel3}). The explicit form in Eq.~(\ref{eq:ModRel3}) for the relations in Eq.~(\ref{eq:Rel3}), which has been presented in Ref.~\cite{Meissner:2007rx} for the first time, looks very similar to the relations of first type in Eq.~(\ref{eq:ModRel1}), but so far it is not known whether Eq.~(\ref{eq:ModRel3}) is restricted to model calculations or whether it could even be valid in general.

Eventually, we were not able to find an explicit form for the relation of fourth type in Eq.~(\ref{eq:Rel4}). Nevertheless, this relation is trivially fulfilled in our model calculations, as the corresponding GPD $\tilde{\mathcal{H}}_T^g$ and TMD $h_{1T}^{\bot g}$ vanish.

\section{\dots and beyond}
So far, from the relations between GPDs and TMDs in Eqs.~(\ref{eq:Rel1})--(\ref{eq:Rel4}) only those of first type in Eq.~(\ref{eq:Rel1}) are known to be valid in general. Therefore, in order to obtain more information on the status of the other types of relations, we analyzed generalized transverse momentum dependent parton distributions (GTMDs). For a spinless target, these are defined through
\begin{equation}
 W^{q[\Gamma]}(x,\xi,\vec{k}_T,\vec{\Delta}_T)
 =\frac{1}{2}\int\frac{dz^-}{2\pi}\,\frac{d^2\vec{z}_T}{(2\pi)^2}\,e^{i k\cdot z}\,
  \big<p'\big|\,\bar\psi\big(\!-\!{\textstyle\frac{1}{2}}z\big)\,\Gamma\,{\mathcal{W}}_{\mathrm{GTMD}}\,
  \psi\big({\textstyle\frac{1}{2}}z\big)\,\big|p\big>\,\Big|_{z^+=0^+} \,,
  \label{eq:DefGTMDs}
\end{equation}
which reduces to the correlator for GPDs in Eq.~(\ref{eq:DefGPDs}) by integration over $\vec{k}_T$ and to the one for TMDs in Eq.~(\ref{eq:DefTMDs}) by setting $\Delta=0$. Note that the correlator in Eq.~(\ref{eq:DefGTMDs}) is directly related to the Wigner distributions discussed in Refs.~\cite{Ji:2003ak, Belitsky:2003nz}.

Using the constraints from hermiticity, parity, and time-reversal, the correlator in Eq.~(\ref{eq:DefGTMDs}) can be parametrized by 16 GTMDs, which are complex-valued functions of $x$, $\xi$, $\vec{k}_T^2$, $\vec{k}\cdot\vec{\Delta}$, and $\vec{\Delta}_T^2$. The four leading twist quark GTMDs of an unpolarized target are
\begin{eqnarray}
 W^{q[\gamma^+]}
 &\!\!\!=\!\!\!&F^q_1(x,\xi,\vec{k}_T^2,\vec{k}_T \cdot \vec{\Delta}_T,\vec{\Delta}_T^2) \,,\\
 W^{q[\gamma^+ \gamma_5]}
 &\!\!\!=\!\!\!&\frac{i\varepsilon_T^{ij} k_T^i \Delta_T^j}{M^2} \,
  \tilde{G}^q_1(x,\xi,\vec{k}_T^2,\vec{k}_T \cdot \vec{\Delta}_T,\vec{\Delta}_T^2) \,,\\
 W^{q[i\sigma^{j+}\gamma_5]}
 &\!\!\!=\!\!\!&\frac{i\varepsilon_T^{ij} k_T^i}{M} \,
  H^{k,q}_1(x,\xi,\vec{k}_T^2,\vec{k}_T \cdot \vec{\Delta}_T,\vec{\Delta}_T^2)
  +\frac{i\varepsilon_T^{ij} \Delta_T^i}{M} \,
  H^{\Delta,q}_1(x,\xi,\vec{k}_T^2,\vec{k}_T \cdot \vec{\Delta}_T,\vec{\Delta}_T^2) \,.\qquad
\end{eqnarray}
From this parametrization one immediately recovers the model-independent validity of the relations of first type in Eq.~(\ref{eq:Rel1}), as the involved GPDs and TMDs are simply limiting cases of the same GTMDs,
\begin{equation}
 \int d^2\vec{b}_T\,{\mathcal{H}}^q(x,\vec{b}_T^{\,2})
 =\int d^2\vec{k}_T\,f_1^q(x,\vec{k}_T^{\,2})
 =\int d^2\vec{k}_T\,{\mathrm{Re}}\bigg[F^q_1(x,0,\vec{k}_T^{\,2},0,0)\bigg] \,.
\end{equation}
For the relations of second type in Eq.~(\ref{eq:Rel2}) one finds, however, that
\begin{equation}
 \Big({\mathcal{E}}_T^{q}+2\tilde{\mathcal{H}}_T^{q}\Big)'
 \sim{\mathrm{Re}}\Big[
  {\textstyle\frac{1}{2}\Big(\frac{k_T^1}{\Delta_T^1}+\frac{k_T^2}{\Delta_T^2}\Big)}\,H^{k,q}_1
  +H^{\Delta,q}_1\Big]
 \qquad{\mathrm{and}}\qquad
 h_1^{\bot q}\sim{\mathrm{Im}}\Big[H_1^{k,q}\Big] \,,
\end{equation}
so that the involved GPDs and TMDs are limiting cases of two independent functions, the real and the imaginary part of some GTMDs. This supports the understanding that the relations in Eq.~(\ref{eq:Rel2}) do not hold in general. At the present stage our analysis does not permit any statement about the relations of third or fourth type in Eqs.~(\ref{eq:Rel3}) and~(\ref{eq:Rel4}), as here we would have to consider, in particular, target polarization.

\section{Conclusions}
We showed that model-independent considerations suggest possible relations between GPDs and TMDs. From these relations, so far only the relations of first type are known to be valid in general. The relations of second type are probably only valid in simple model calculations, which is supported by our analysis of GTMDs. It will be very interesting to redo this analysis for the relations of third and fourth type, as at least the relations of third type are similar to those of first type and could therefore be valid in general.

\section*{Acknowledgments}
This work has partially been supported by the Verbundforschung ``Hadronen und Kerne'' of the BMBF and by the Deutsche Forschungsgemeinschaft (DFG).\\[2.5mm]
{\bfseries Notice:} Authored by Jefferson Science Associates, LLC under U.S. DOE Contract No. DE-AC05-06OR23177. The U.S. Government retains a non-exclusive, paid-up, irrevocable, world-wide license to publish or reproduce this manuscript for U.S. Government purposes.



\begin{thebibliography}{99}
 \bibitem{Burkardt:2002ks:2003uw}
  M.~Burkardt,
  Phys.\ Rev.\  D {\bf 66}, 114005 (2002);
  Nucl.\ Phys.\  A {\bf 735}, 185 (2004).
 \bibitem{Burkardt:2003je}
  M.~Burkardt and D.~S.~Hwang,
  Phys.\ Rev.\  D {\bf 69}, 074032 (2004).
 \bibitem{Diehl:2005jf}
  M.~Diehl and P.~H{\"a}gler,
  Eur.\ Phys.\ J.\  C {\bf 44}, 87 (2005).
 \bibitem{Burkardt:2005hp}
  M.~Burkardt,
  Phys.\ Rev.\  D {\bf 72}, 094020 (2005).
 \bibitem{Lu:2006kt}
  Z.~Lu and I.~Schmidt,
  Phys.\ Rev.\  D {\bf 75}, 073008 (2007).
 \bibitem{Meissner:2007rx}
  S.~Meissner, A.~Metz, and K.~Goeke,
  Phys.\ Rev.\  D {\bf 76}, 034002 (2007).
 \bibitem{Ji:2003ak}
  X.~d.~Ji,
  Phys.\ Rev.\ Lett.\  {\bf 91}, 062001 (2003).
 \bibitem{Belitsky:2003nz}
  A.~V.~Belitsky, X.~d.~Ji, and F.~Yuan,
  Phys.\ Rev.\  D {\bf 69}, 074014 (2004).
\end{thebibliography}
\end{document}